\documentclass[12pt]{article}

\usepackage{fullpage, doublespace}
\newcommand{\Cpp}{C$^{\mbox{{\tiny++}}}$}

\begin{document}
\title{Influencing Software Usage}

\author{\begin{tabular}{ccc}
Lorrie Faith Cranor & & Rebecca N. Wright\\
\end{tabular} \\ \\
AT\&T Labs --- Research\\
{\sl \{lorrie, rwright\}@research.att.com}}

\date{}

\begin{singlespace} 
\maketitle

\begin{abstract}

Technology designers often strive to design systems that are flexible
enough to be used in a wide range of situations. Software engineers,
in particular, are trained to seek general solutions to
problems. General solutions can be used not only to address the
problem at hand, but also to address a wide range of problems that the
designers may not have even anticipated. Sometimes designers wish to
provide general solutions, while encouraging certain uses of their
technology and discouraging or precluding others. They may attempt to
influence the use of technology by ``hard-wiring'' it so that it only
can be used in certain ways, licensing it so that those who use it are
legally obligated to use it in certain ways, issuing guidelines for
how it should be used, or providing resources that make it easier to
use the technology as the designers intended than to use it in any
other way. 

This paper examines several cases where designers have attempted to
influence the use of technology through one of these mechanisms.  Such
cases include key recovery encryption, Pegasus Mail, Platform for
Internet Content Selection (PICS) Guidelines, Java, Platform for
Privacy Preferences Project (P3P) Implementation Guide, Apple's style
guidelines, and Microsoft Foundation Classes.  In some of these cases,
the designers sought to influence the use of technology for
competitive reasons or in order to promote standardization or
interoperability. However, in other cases designers were motivated by
policy-related goals such as protecting privacy or free speech. As new
technologies are introduced with the express purpose of advancing
policy-related goals (for example, PICS and P3P), it is especially
important to understand the roles designers might play in influencing
the use of technology.

\end{abstract}
\end{singlespace}


\section{Introduction} \label{sec-intro}

Technology designers often strive to design systems that are flexible
enough to be used in a wide range of situations. Software engineers,
in particular, are trained to seek general solutions to problems.
General solutions can be used not only to address the problem at hand,
but also to address a wide range of problems that the designers may
not have even anticipated. For example, spreadsheet programs often
come with templates for computing interest payments, keeping track of
personal finances, and performing other operations that their
designers consider useful. But users can also use these programs to
record and analyze virtually any kind of tabular data. Thus these
programs are not limited by the ability of their designers to
anticipate ways that spreadsheets will be used. Indeed, users can and
do invent their own spreadsheet applications.

While general technology solutions allow users to use the technology
as they see fit, users may not always use the technology in desirable
ways. Users may use general-purpose technology tools in ways that are
inefficient, dangerous, or contrary to the spirit in which the
technology was designed. 

The Hypertext Transfer Protocol (HTTP) Referer field provides an
example of the danger of using technology in an unintended way. HTTP
is the protocol used for requesting and transferring Web pages over
the Internet~\cite{HTTP10}. The Referer field is an optional HTTP
header that allows Web site operators to track the addresses of Web
pages from which visitors to their site are referred. However, some
Web sites use the Referer field for another purpose for which it was
never intended: to verify that forms submitted by visitors have not
been improperly modified. This is a very weak authentication mechanism
that can be defeated easily by forging the Referer field's
contents. Furthermore, as the Referer field is an optional header,
there is no guarantee that Web browsers will transmit it to a Web
site. Indeed, to address a security problem, Microsoft modified the
behavior of the Internet Explorer browser beginning with version 4.0
so that the Referer field is not transmitted under certain
conditions~\cite{mkbq178066}. As a result, some sites that use the
Referer field in this unintended way are finding that users are unable
to submit forms at their site. In a column describing this problem,
Lincoln D. Stein concluded, ``This painful lesson illustrates why it's
almost never a good idea to use a feature for other than its intended
purpose''~\cite{stein98}.

Some unintended usage of technology is neither inefficient nor
dangerous, but it nonetheless violates the spirit in which the
technology was designed. For example, some email programs are
efficient tools for sending out bulk email messages. One such tool,
Pegasus Mail, was developed by programmer David Harris to provide a
free service ``that enhances the quality of communication between
people''~\cite{harris98-terms}. However, many people have used Pegasus
mail to send unsolicited commercial email, better known as ``spam.''
Spam is often blamed for interfering with interpersonal communication,
increasing the time and effort necessary to retrieve and identify
desired email~\cite{cranor98-spam}. Thus when Pegasus Mail is used for
sending spam it violates the spirit in which this technology was
designed.

Sometimes designers wish to provide general solutions, while
encouraging certain uses of their technology and discouraging or
precluding others. In this paper we identify four types of mechanisms
used by software engineers (or their employers) to influence the use
of technology. Software engineers may attempt to influence the use of
technology by: 
\begin{itemize}
\item ``hard-wiring'' it so that it only can be used in certain ways,
\item licensing it so that those who use it are legally obligated to use it
in certain ways, 
\item issuing guidelines for how it should be used, or
\item providing resources that make it easier to use the technology as
the designers intended than to use it in any other way.
\end{itemize}
We provide several examples of software systems to which each of these
mechanisms has been applied. In some of these cases, the designers
sought to influence the use of technology for competitive reasons or
in order to promote standardization or interoperability. However, in
other cases designers were motivated by policy-related goals such as
protecting privacy or free speech.  In some cases, the designers did
not actually seek to influence the use of technology, but their
actions had inadvertent influence.

As new technologies are introduced with the express purpose of
advancing policy-related goals (for example, PICS and P3P), it is
especially important to understand the roles designers might play in
influencing the use of technology.

\section{Hard-Wiring} \label{sec-hard-wiring}

By ``hard-wiring'' software systems, designers can ensure that
software behaves in a certain way and thus influence how it can be
used.

\subsection{Key Recovery Encryption}\label{subsec-recovery}

Key recovery encryption, also sometimes called key escrow encryption
or trusted third-party encryption, is the term for encryption schemes
in which the secret encryption keys may be recovered by certain
parties in certain circumstances.

The U.S. government advocates key recovery encryption as providing a
balance between the need on the one hand for individual privacy and
legitimate commercial activities, and on the other hand the need for
government eavesdropping for purposes of law enforcement or national
security~\cite{s909}.  However, not everyone agrees that these
concerns necessarily conflict, nor that such a balance is best
provided by key recovery encryption~\cite{AAB+98}.

The U.S. government claims any key recovery encryption system will be
voluntary to use~\cite{admin}, and that citizens will be free to use
whatever encryption software they wish (though some also fear that
this is merely a stepping stone to disallowing all non-recoverable
encryption systems, as has already been done in countries such as
France, Russia, and Indonesia).  However, the government hopes that
key recovery encryption will become widely used by hard-wiring it into
standardized, widely available solutions, such as secure telephones,
Web browsers, and a global public key infrastructure~\cite{rules}.
Furthermore, the government can further propagate the use of its
sanctioned key recovery encryption solutions by allowing them to be
exported freely to other countries, while strong encryption solutions
that do not allow key recovery are currently heavily restricted
against export~\cite{rules}.

Indeed, despite industry's initial opposition to key recovery
encryption, several companies have realized that because of the
government's willingness to provide export licenses for key recovery
encryption and not to routinely provide them for products including
non-recoverable encryption, it may be commercially more viable to
support key recovery.  To this end, over 30 companies from around the
world have formed the ``Key Recovery Alliance'' (KRA) to ``promote the
implementation, deployment, and use of interoperable key recovery
technologies in the global marketplace.''~\cite{kra}.  KRA members
include Apple, IBM, Digital Equipment Corporation, Hewlett-Packard,
and Verisign.  In addition, other companies, such as Netscape, have
agreed to deploy key recovery encryption in order to obtain export
licenses for their products.  As more businesses routinely include key
recovery encryption in their products, more users will use key
recovery encryption even if they are unaware that this is the case;
this is discussed more in Section~\ref{subsec-recovery-toolkits}.

\subsection{Pegasus Mail}\label{subsec-pegasus-wiring} 

In order to prevent users of Pegasus Mail from using it to send spam,
Harris might have hard-wired Pegasus Mail so that it could not be used
in this way.  However, this would have likely reduced the program's
utility as a communication tool, making it difficult or impossible for
people to use it to send legitimate bulk email. Harris took a
different approach and modified Pegasus Mail (beginning with version
2.50) so that it automatically inserts ``X-Distribution'' headers into
all email messages it sends that are distributed to 50 or more
recipients. Harris also distributed instructions for filtering spam
identified by the presence of these headers. Harris explained that
this step should reduce ``the abuse of Pegasus Mail ... yet it does
not significantly impact on legitimate users of the
program''~\cite{Harris97}.  As discussed in
\ref{subsec-pegasus-licensing}, Harris also uses licensing to control
the use of Pegasus Mail for sending spam.

\subsection{PICS Implementations}\label{subsec-pics-implementation}
The Platform for Internet Content Selection (PICS) is a series of
specifications developed by the World Wide Web Consortium (W3C) to
provide a standard platform for rating Internet content. PICS allows
anyone to come up with a set of rating criteria---called a rating
system---and use it to rate Web pages. The development of PICS was
largely motivated by a desire to give parents tools they could use to
protect their children from harmful online materials (as an
alternative to laws that would prohibit the online distribution of
materials harmful to children). However, PICS was designed to be a
more general specification for content rating that would be useful for
a wide variety of applications beyond protecting
children~\cite{resnick96}. But despite the flexibility of the PICS
specifications, most PICS implementations to date have hard-wired
settings that make them ill-suited for many applications that go
beyond protecting children.  The Netscape Navigator 4.5 implementation
even goes so far as to hard-wire in two specific PICS rating systems.

Content Advisor---the PICS implementation in the Microsoft Internet
Explorer Web browser---has a user interface that will read content
labels from any PICS rating system, but can only be configured to
select content with numeric ratings \emph{less than} a specified
threshold. This makes sense for some of the most common PICS ratings
systems that rate characteristics such as `nudity' and `violence' on
increasing numeric scales. A parent who wants to prevent her child
from viewing content with a violence rating higher than 3 simply
positions the slider on the Content Advisor interface so that only
content with violence rating 3 or lower is permitted. The interface is
hard-wired so that it is impossible to configure it to accept only
content rated as having \emph{high} levels of nudity or violence (or
any other criteria) or to select ratings in the middle of a
range~\cite{cranor98}. It is not clear whether the interface designers
restricted the interface so as to thwart efforts by children to seek
out sexual or violent content, or whether the interface was designed
this way for other reasons (or even arbitrary reasons). However, it
has the effect of limiting the ways that Content Advisor can be
used. A likely unintended consequence is that the interface cannot be
used to filter out low quality content using a PICS rating system that
assigns numeric ratings that correspond to quality stars (four stars
for high quality content, zero stars for low quality content, etc.).

\subsection{Default Settings} \label{subsec-defaults} 

Even when they are not ``hard-wired'' and may even be relatively easy
to change, the settings initially provided by the manufactures are
often left untouched by users.  Consequently, the default settings
that inevitably accompany most pieces of software often have great
influence on how that software is used. In some cases, designers are
cognizant of this and design the default settings with an eye towards
influencing user behavior. In other cases, default settings are based
on somewhat arbitrary decisions, but they may be nonetheless
influential. There is evidence that users rarely change default
settings in popular consumer software products. For example, Netscape
reports that 40 percent of the visitors to their Web site come there
because they have not changed their default home
page~\cite{seidman98}.

When designing software that includes features that can be configured
in multiple ways, designers are often faced with a choice of (a)
configuring the features with default settings, (b) leaving the
features that require configuration turned off, or (c) leaving
features that require configuration unconfigured, but prompting users
to configure them before they can use the
software~\cite{cranor97}. While explicitly configuring features with
default settings is an obvious way to influence software use, leaving
features that require configuration turned off can also be influential
as it limits the degree to which those features will likely be used at
all.

\section{Licensing} \label{sec-licensing}

Software may be licensed so that those who use it are legally
obligated to use it in certain ways.

\subsection{Pegasus Mail}\label{subsec-pegasus-licensing} In addition
to hard-wiring Pegasus Mail to insert special headers when the program
is used to send bulk email as described in
Section~\ref{subsec-pegasus-wiring}, Harris also wrote a prohibition
against spamming into the Pegasus Mail license. The Pegasus Mail Terms
and Conditions of Use specifically prohibits using Pegasus Mail for
sending email to ``more than 50 recipients for the purpose of
advertising a commercial product or service, where the recipient has
not explicitly expressed interest in receiving such advertisements.''
The license also prohibits including the program in a package designed
for sending spam or promoting the use of the program for that
purpose~\cite{harris98-terms}.  However, Harris speculates that the
cost of enforcing this license in the U.S. legal system would likely
be prohibitive for him~\cite{harris98}.  Indeed while the threat that
licenses may be enforced may be somewhat effective in deterring
unauthorized use, the ultimate effectiveness of licenses is likely to
depend on their ability to be enforced.

\subsection{Java}\label{subsec-java} Sun Microsystems' Java technology
is designed to allow software developers to create ``write-once,
run-anywhere'' code.  That is, any code written in Java should run on
any computer, or in any Web browser, that supports Java, without the
developer having to know anything about which computers or browsers
the code will be run on.  The Java technology license carries two
major restrictions on its use.  The first places restrictions on its
use for critical applications that it may not be suitable for, while
the second ensures compatibility among different Java language based
products, thereby maintaining the write-once, run-anywhere philosophy.

More specifically, the first restriction, found in the Java
Developer's Kit (JDK) on-line binary code license agreement applies to
anyone developing Java applications, and states that the ``software is
not designed or licensed for use in on-line control equipment in
hazardous environments such as operation of nuclear facilities,
aircraft navigation or control, or direct life support
machines''~\cite{jdk-binary}. While this may be hard to enforce due to
the problem of detection before problems occur, it makes clear Sun's
impunity to lawsuits in the event that such use of Java results in
damages.

The second restriction is only required for a commercial source
license, needed if the user intends to merge the Java source code into
a commercial product, such as a commercial Web browser that supports
the Java interpreter and class libraries.  The commercial source code
license is devoted to ensuring cross-platform Java compatibility.
Compatibility requirements are ensured partly by restrictions on the
kinds of changes that can be made to the underlying Java technology
components provided by Sun, and are tested through a detailed series
of ``test suites.'' For example, the commercial source license between
Sun and Microsoft includes the following~\cite{jdk-sources}:
 
\begin{quotation}
Licensee agrees that upon \dots (a) six (6) months after the date that
SUN delivers to Licensee an Upgrade that SUN designates a significant
Upgrade (each, a ``Significant Upgrade''), \dots Licensee shall
deliver to SUN \dots an upgrade to the Java Reference Implementation
(each, a ``Compatible Implementation'') that passes the test suite
that accompanied the Significant upgrade (a ``Relevant Test
Suite'')\dots 

Licensee shall confine the names of all VAOPs [Value Added Open
Packages] to names beginning with ``COM.ms'' and shall not modify or
extend the names of public class or interface declarations whose names
begin with ``java'', ``COM.sun'' or their equivalents\dots 

Licensee agrees that the Reference Implementation VM [Virtual
Machine], and any upgrades thereto, shall include the necessary Source
Code to implement the functionality of the Java Runtime Interpreter.
Notwithstanding the foregoing, Licensee may implement a subset of the
functionality\dots callable by a system-level interface (a ``System
Implementation'').  In such cases, Licensee shall provide SUN a
complete and accurate specification for all such system-level
interfaces\dots

Licensee agrees that the Reference Implementation Java Classes shall
be maintained as an application level library.
\end{quotation}

The license also says that any Java classes added by the licensee to
the existing set of Java classes should be made promptly available to
developers and customers free of charge, and that for any ``Portable
Value Added Open Packages,'' either the source code or a complete and
accurate specification including an appropriate test suite must be
made publicly available.

Because Sun views cross-platform compatibility as a fundamentally
important aspect of Java, it has been aggressive about enforcing such
requirements.  Indeed, Sun sued Microsoft in October of 1997 for
alleged breaches of their license in Microsoft's Internet Explorer 4.0
resulting in incompatibilities with the Java standard.  At the time of
this writing, the suit still continues, with hearings scheduled in
September 1998.

\subsection{PICS}\label{subsec-pics} Licensing has been considered and
rejected for some technologies due in part to the difficulty of
defining the scope of a license and following up with enforcement. For
example, PICS was intended to be an alternative to legislative
proposals that would ban or restrict certain kinds of content; PICS
was designed to be a voluntary system that would let parents decide
for themselves what content is appropriate for their
children~\cite{resnick96}. However, PICS drew criticism from free
speech advocates who were concerned that one or two PICS rating
systems could become de facto standards and eventually become subject
to legislation requiring their use~\cite{aclu97}. In response to
frequent questions about why the PICS designers had not licensed PICS
so as to preclude its being used in ways that might be harmful to free
speech, one of the PICS designers responded that at the time PICS was
developed licensing was viewed as infeasible because it would
``undercut the `neutrality' and appeal of the technology'' and because
``the W3C then would have had to be in the position of determining who
should and should not use it; this is not a role the W3C is competent
to play''~\cite{resnick98}.

\section{Guidelines} \label{sec-guidelines} 

Designers sometimes issue guidelines about how software or systems
should be used. Guidelines do not limit the ability of technology to
be used in unexpected ways, but they can provide guidance as to the
way technology should be used. Guidelines can also be issued after
software is released and they can be updated as people gain experience
using a technology.

While guidelines can make explicit the intention of designers and
provide guidance to those who care to take advantage of them, they are
not requirements, and they often can be overlooked or ignored. The
impact of guidelines may be increased by logo or seal programs that
provide a logo or seal of approval that products following a
particular set of guidelines are allowed to display. However, these
programs generally require an arbiter to determine which products are
entitled to display the logo, and permission to display the logo may
be tied to a licensing agreement.

There may also be a risk that users (as well as implementors) may
confuse guidelines with {\em requirements\/}, and come to expect that
products follow published guidelines. Thus users may develop false
expectations about products, or implementors may build interoperable
products that interoperate only with products following a particular
set of guidelines.  On the other hand, if users or some implementors
treat guidelines as requirements, it may put more pressure on other
implementors to follow the guidelines.

\subsection{PICS}\label{subsec-pics-guidelines}

Over two years after PICS was developed, the PICS designers issued a
``Statement on the Intent and Use of PICS: Using PICS Well.'' This
document includes a set of guidelines that address how the designers
intended PICS to be used~\cite{pics98}. It's not clear that PICS had
actually been used in an undesirable way, but criticism that PICS had
been designed such that it could be used in undesirable ways
(especially from a first amendment standpoint) likely prompted the
PICS designers to issue this statement~\cite{aclu97,cranor98}.

\subsection{P3P}\label{subsec-p3p} Another W3C specification, the
Platform for Privacy Preferences Project (P3P)~\cite{reagle98},
developed guidelines in parallel with the specification
itself. Motivated by both the response to PICS and a perceived need to
provide guidance on issues outside the scope of the P3P specification
(such as user interface), the P3P designers began developing an
implementation guide. One section of the guide, the P3P Guiding
Principles~\cite{cranor98b}, provides a set of guidelines for the way
P3P should be implemented and used.  These guidelines emphasize that
P3P was designed as a privacy protection tool and advise implementors
to make design decisions that support the goals of P3P. However, the
document notes that the guidelines are recommendations rather than
requirements.


\subsection{Apple Style Guidelines}\label{subsec-apple-guidelines}

Apple Computers has a deeply thought-out philosophy behind the design
of its computers and their software, reflected in almost every detail,
from the use of color to the placement of menus.  In order to help
ensure a ``look and feel'' consistent with this philosophy across all
software applications that run on the Macintosh, Apple Computers has
provided an extensive list of ``style guidelines'' to be used by
developers of software applications for the Macintosh.  Specifically,
their 400-page book {\em Macintosh Human Interface
Guidelines}~\cite{mac-hci-guide} ``describes the way to create
products that optimize the interactions between people and Macintosh
computers.  It explains the whys and hows of the Macintosh interface
in general terms and specific details.''~\cite{mac-hci-guide}.  This
book is comprehensive and easy-to-read, with lots of examples of
``good'' and ``bad'' software development for the Macintosh; it is
available both for free in electronic format from the Apple Web site,
and in print from conventional book-sellers.  In addition, other style
guidelines documents such as ``Control Layout Guidelines,''
``Guidelines for Content Manipulation,'' and ``Application Layout
Guidelines'' are available electronically from their Web site.  In
making such information available and readable, Apple has made it much
easier for developers to consistently design products that fit the
Macintosh model and philosophy, and many developers have indeed done
so.  The guidelines are also supported by a toolkit designed to help
developers follow the guidelines; this is discussed below in
Section~\ref{subsec-apple-toolkit}.

\section{Toolkits and other Resources} \label{sec-toolkits} 

Often, the designers of software and computer systems provide
resources that make it easier to use the technology as the designers
intended than to use it in any other way.  Such resources can be in
the form of toolkits, published interface specifications, or even
simply by making certain off-the-shelf solutions widely available.

\subsection{Key Recovery Encryption}\label{subsec-recovery-toolkits}
As an example of influencing usage via off-the-shelf availability, the
U.S. government may succeed in its goal of making most encryption that
people use recoverable simply by getting most commercial software
developers to include it hard-wired in their systems as discussed in
Section~\ref{subsec-recovery}.  Similarly, if a global public key
infrastructure using key recovery is put into place, the comparative
ease of using it rather than having to do encryption without the
benefit of a public key infrastructure may go a long way towards
making most encrypted traffic recoverable.

On the other hand, people who don't like key recovery can try to
combat this by making widely available non-recoverable encryption
solutions.  For example, until its recent acquisition by Network
Associates, the PGP (Pretty Good Privacy) encryption
system~\cite{zim95} specifically did not allow key recovery, was
freely available on the Internet for personal, noncommercial use, and
built up a ``grass-roots'' public key infrastructure in which
individual users certify each other's public keys.  PGP's creator,
Phil Zimmerman, also used PGP as a staging ground for debates on
individuals' rights to privacy, cryptography export regulations, and
certain patentability issues.

\subsection{Apple Style Guidelines}\label{subsec-apple-toolkit} As
described in Section~\ref{subsec-apple-guidelines}, Apple has
extensive guidelines describing what it considers to be good software
development practices for the Macintosh.  To further encourage
developers to follow their guidelines, they also provide a toolkit of
Macintosh system software routines that follow the guidelines and can
be used by developers in their Macintosh software
applications~\cite{mac-toolkit}.  Since it is often easier to use the
toolkit routines than to write similar routines from scratch, this
results in different developers using the same toolkit routines,
thereby creating a more consistent and standard interface to be
presented to the user even when using software from different
developers.

\subsection{Microsoft Foundation Classes}\label{subsec-mfc}

The Microsoft Foundation Class Library (MFC) is a software library
consisting of a ``set of reusable, pre-built \Cpp\ components that
provide a completely portable interface for applications for the
Microsoft Windows, Windows 95, and Windows NT operating systems,'' as
well as some Macintosh and Unix platforms~\cite{msft}.  Its main goal
is to allow developers to write applications for Windows much faster
than writing them from scratch.  This is an important marketing
feature for Microsoft because it results in more Windows applications
being available even as new versions of Windows are constantly
evolving.  Although the MFC to some extent results in a similar look
and feel across different applications, this does not appear to have
been as important a consideration for Microsoft as it was for Apple
(see Sections~\ref{subsec-apple-guidelines}
and~\ref{subsec-apple-toolkit}).  Instead, they have chosen to make
the MFC quite general to allow developers to have more flexibility
even though this tends to result in a more diverse look and feel
across different applications. 

In addition to the MFC itself, Microsoft has also provided a set of
guidelines for developers who want to extend the MFC in a way that
will interact well with existing MFC components~\cite{msft-guide}.

\section{Conclusions} \label{sec-conclusion}

We have discussed four different types of methods that software
engineers may use to attempt to influence the way their technology is
used. We have provided a variety of examples of how these techniques
have been used, with varying degrees of success. From our analysis,
there is no one method that emerges as most successful. 
\begin{itemize}
\item Hard-wiring certainly is effective in limiting the use of a
particular tool, but it may prove overly limiting, precluding many
desirable (and possibly unforeseeable) uses of that tool. Furthermore,
there may be alternative tools, possibly from competing vendors,
readily available to those who are determined to use the technology in
ways that designers are attempting to influence against.
\item Licensing can be effective when licenses are readily
enforceable. However, enforcing licenses can be expensive and
time-consuming, both due to the difficulty of monitoring licensees to
determine when licenses are not being enforced, and due to the legal
expenses involved in pursuing a case against a misbehaving licensee.
While the threat of enforcement may be somewhat of a deterrent,
undesirable use is not actually precluded, although it may sometimes
be punished after it occurs.  Licensing can also be useful to limit
the liability that a software developer has if their technology is
used in unapproved ways; this may exert further influence on users to
follow approved usage.
\item Guidelines have the advantages that they do not limit unexpected
uses, and they can be updated as people gain experience with a new
technology. However, they neither preclude nor punish undesirable
use. Logo programs and pressure from users or implementors may
encourage voluntary compliance.
\item Toolkits and other resources can make it much easier for people
to use technology as intended than to use it in other ways that will
require them to build the necessary resources themselves. However,
when people feel strongly about using technology in ways not supported
by the toolkit or other available resources, they can and will develop
their own resources that result in uses that designers are trying to
influence against.
\end{itemize}

Designers who wish to influence software usage would be wise to
consider the tradeoffs of each of these methods when determining what
approach to take in a particular situation.  However, designers should
also keep in mind that while these methods can be useful in persuading
users who are choosing between multiple alternatives they consider
acceptable, these methods are unlikely to persuade users to choose
alternatives to which they are opposed.

\bibliography{tprc}

\end{document}